\documentclass[conference]{IEEEtran}
\IEEEoverridecommandlockouts
\usepackage{subfigure}
\usepackage{verbatim}
\usepackage{amsmath}
\usepackage{cite}
\usepackage{amsmath,amssymb,amsfonts}
\usepackage{algorithmic}
\usepackage{graphicx}
\usepackage{textcomp}
\usepackage{xcolor}
\def\BibTeX{{\rm B\kern-.05em{\sc i\kern-.025em b}\kern-.08em
    T\kern-.1667em\lower.7ex\hbox{E}\kern-.125emX}}
\begin{document}

\title{Finite-time Motion Planning of Multi-agent Systems with Collision Avoidance}

\author{\IEEEauthorblockN{Yilei Jiang}
\IEEEauthorblockA{\textit{Department of Computer Science} \\
\textit{The Chinese University of Hong Kong}\\
HKSAR, China \\
yljiang@link.cuhk.edu.hk}
\and
\IEEEauthorblockN{Dongkun Han}
\IEEEauthorblockA{\textit{Department of Mechanical and Automation Engineering} \\
\textit{The Chinese University of Hong Kong}\\
HKSAR, China \\
dkhan@mae.cuhk.edu.hk}}
\maketitle

\begin{abstract}
Finite-time motion planning with collision avoidance is a challenging issue in multi-agent systems. This paper proposes a novel distributed controller based on a new Lyapunov barrier function which guarantees finite-time stability for multi-agent systems without collisions. First, the problem of finite-time motion planning of multi-agent systems is formulated. Then, a novel finite-time distributed controller is developed based on a Lyapunov barrier function. Finally, numerical simulations demonstrate the effectiveness of proposed method.
\end{abstract}


\section{Introduction}
Multi-agent systems have received much interest in recent years, thanks to its applicability to a wide range of study fields and real-world applications. Various distributed coordination and control problems, such as consensus (also known as agreement/synchronization/rendezvous), formation, distributed observation, estimation and optimization, have been introduced and implemented in many practical industrial applications \cite{pecora90prl,jlu04cas1,zli10}. \\

When it comes to multi-vehicle systems in particular, a common ground may be the requirement for several agents to collaborate in order to achieve one or more cooperative goals. In such circumstances, the available patterns on sensing and information exchange, as well as physical/environmental limits and intrinsic limitations, inherently dictate coordination and control. As a result, motion planning, coordination, and control has been, and continues to be, a hot focus of research in the robotics and control societies. \\

Inter-agent collision avoidance, convergence to spatial destinations/regions or tracking of reference signals/trajectories, maintenance of information exchange among agents, and avoidance of physical obstacles are the main concerns when coordinating the motions of multi-vehicle or multi-robot (the terms are used interchangeably) teams. \cite{b1,b2,b3,b4,b5,b6} deal with flocking and \cite{b7,b8,b9,b10,b11,b12} works on consensus, rendezvous, and/or formation control. Collision avoidance is a non-negotiable condition in such problems, and potential function approaches and Lyapunov-based analysis provide useful and effective ways to handle it. Recently, a method based on Lyapunov barrier function is proposed and shows its soundness in coping with multi-task multi-agent problems \cite{panagou16tac, han19tac}. Please check the survey of formation control by using artificial potential functions with various different setting-ups \cite{b13}.\\

Despite these objectives, finite time stability is, however,
seldom achieved. Finite Time Stability (FTS) is of essential importance because in real cases multiple
agents need to one or multiple common goals within limited time. [14] focuses on
continuous autonomous systems and provides necessary and sufficient conditions based on the Lyapunov stability theory. In \cite{b15,b16,b17,b18}, consensus and formation control problems within finite
time framework are well investigated by several classes of protocols. An inspiring work studied the problem of FTS for a single system by using barrier functions in \cite{b19}. \\

Different from the work \cite{b19}, we propose a finite-time distributed controller without
collision for multi-agent system, and the contribution of this paper is threefold. First, a new type of Lyapunov barrier function is constructed, whose derivative satisfy specific properties such that its gradient is non-zero everywhere except the equilibrium point. Second, by using the developed Lyapunov barrier function, we provide the distributed controller for multi-agent systems with considering collision avoidance. Finally, distinct with the work in literature, the benefits of proposed method can also be found in the situations where unsafe region or boundaries of safe region exists, which allows a more practical implementation environment for multi-agent systems. \\

\section{Modeling and Problem Statement}
Consider a network of $N$ mobile agents deployed in a known workspace $W$. Each agent $i \in\{1, \ldots, N\}$ is modeled as a circular disk of radius $r$ and circular agents are centered at known positions $\boldsymbol{x}_{\boldsymbol{i}}$, $i \in\{1, \ldots, N\} .$ Moreover, we assume all kinetic agents are located within a bounded circle of radius $R$. To achieve this, we place M static agents with radius of $r$ along the circular boundary. With collision avoidance of agents on the boundary, kinetic agents remain in the circle according to designed barrier function. As a result, network connectivity is preserved and one agent can receive the position and velocity information of other agents. Now consider one kinetic agent $i$. Its motion under single integrator dynamics is
\begin{equation}
\dot{\boldsymbol{x}}_{\boldsymbol{i}}=\boldsymbol{v}_{\boldsymbol{i}}
\end{equation}

where \[\boldsymbol{x}_{\boldsymbol{i}}, \boldsymbol{v}_{\boldsymbol{i}} \in \mathbf{R}^{\mathbf{n}} \]. The problem of reaching to a specified goal position in finite time can be formulated mathematically as follows:
\begin{equation}
\exists t^{*}<\infty s . t \forall t>t^{*}\left\|\boldsymbol{x}_{i}(t)-\boldsymbol{\tau}_{\boldsymbol{i}}\right\|=0    
\end{equation}
where $\tau_{i}$ is the desired goal location. Collision avoidance between agents can be written as:
\begin{equation}
\forall t>t_{0}\left\|\boldsymbol{x}_{\boldsymbol{i}}(t)-\boldsymbol{x}_{\boldsymbol{j}}(t)\right\|>d_{c}
\end{equation}

where $\boldsymbol{x}_{\boldsymbol{i}}(t)$ and $\boldsymbol{x}_{\boldsymbol{j}}(t)$ represent the location of the agents and $t_{0}$ is the starting time. Here,we assume that $d_{c}$ is bigger than $2 r .$ and the distances between goal positions $\left\|\boldsymbol{\tau}_{\boldsymbol{i}}-\boldsymbol{\tau}_{\boldsymbol{j}}\right\|>d_{c} .$ We also assume that one agent starts sufficiently far away from the other agents so that $\left\|\boldsymbol{x}_{\boldsymbol{i}}\left(t_{0}\right)-\boldsymbol{x}_{\boldsymbol{j}}\left(t_{0}\right)\right\|>d_{c} .$ What we should do is to design a feedback-law $\boldsymbol{v}_{\boldsymbol{i}}$ such that all agents reach their own goal positions within finite time while maintaining safe distance between each other.

\section{Motion Coordination}
All the agents initiate in the closed region where the wireless communication links can be established. As the agents are moving inside the region during the process, their wireless communication can remain stable, thus resulting in connectivity maintenance. For any one of the agents, for example the $i^{t h}$ agent $x_{i}$, can not only know the location of the boundary but also perceive the position and velocity of its nearest agent, for example $x_{j}$. We assume that:
(1) The goal positions $\tau_{i}, i \in\{1,2, \ldots, n\}$ are static;
(2) There are no physical obstacles in the region;
(3) The distance between any two agents' goal positions $\left\|\tau_{i}-\tau_{j}\right\|>d_{c}$;
(4) All agents are equal, i.e., there is no leader among them.
Since all the agents are equal, it suffices to focus on one agent. We seek a continuous feedback law $\boldsymbol{u}_{i}$ for the $i^{t h}$ agent to achieve multi-agent coordination with obstacle avoidance. More specifically, we seek a barrier-function based controller for the system. We define the barrier function for the $i^{\text {th }}$ agent as follows:
\begin{equation}
B_{i}\left(\boldsymbol{x}_{\boldsymbol{i}}, \boldsymbol{x}_{j}\right)=\frac{\left\|\boldsymbol{x}_{i}-\tau_{i}\right\|^{2}}{\left\|\boldsymbol{x}_{i}-\boldsymbol{x}_{j}\right\|-d_{c}+\frac{1}{\epsilon}}
\end{equation}

where $\epsilon \gg 1$ is a very large number. We define the controller as follows:
\begin{equation}
v_{i}= \begin{cases}-k_{1}\left\|\nabla B_{i x_{i}}\right\|^{\alpha-1} \nabla B_{i x_{i}}+\left(1-\frac{2\left(x_{i}-\tau_{i}\right)^{\mathrm{T}} \cdot v_{j}}{x_{0}\left(\nabla B_{i x_{i}}\right)^{\mathrm{T}} \cdot v_{j}}\right) v_{j} \\ \hfill \quad \boldsymbol{x} \neq \tau_{i} \\ 0 \hfill \quad  \boldsymbol{x}=\tau_{i}\end{cases}
\end{equation}

where $k_{1}>0$ and $0<\alpha<1$
With this controller, we have the following result:\\ \\ \hspace*{\fill} \\
$\boldsymbol{{Theorem 1: }}$ Under the control law, the point $$\boldsymbol{x}=\boldsymbol{\tau}_{\boldsymbol{i}}$$ is FTS equilibrium for the system and t.  he agent will remain collision avoidance w.r.t any other agents. \\

Before presenting the proof, we present some useful Lemmas:\\ 
Lemma 1 :Under the control law, the point $\boldsymbol{x}=\boldsymbol{\tau}_{\boldsymbol{i}}$ is an equilibrium for the system, i.e.,
\begin{multline}
\lim _{x_{i}  \rightarrow \tau_{i}}-k_{1}\left\|\nabla B_{i x_{i}}\right\|^{\alpha-1} \nabla B_{i x_{i}}+\\
\left(1-\frac{2\left(\boldsymbol{x}_{i}-\tau_{i}\right)^{\mathrm{T}} \cdot \boldsymbol{v}_{j}}{x_{0}\left(\nabla B_{i x_{i}}\right)^{\mathrm{T}} \cdot \boldsymbol{v}_{j}}\right) \boldsymbol{v}_{\boldsymbol{j}}=\mathbf{0}
\end{multline}

Proof: Consider:\\
\begin{align}
\frac{\left(\nabla B_{i x_{i}}\right)^{\mathrm{T}} \cdot v_{j}}{\left(x_{i}-\tau_{i}\right)^{\mathrm{T}} \cdot v_{j}}&=\frac{\left[2\left(\boldsymbol{x}_{i}-\tau_{i}\right)^{\mathrm{T}}-\frac{\left\|\boldsymbol{x}_{i}-\tau_{i}\right\|^{2}}{x_{0}}\left(\boldsymbol{x}_{i}-\boldsymbol{x}_{j}\right)^{\mathrm{T}}\right] \cdot \boldsymbol{v}_{j}}{\left(\boldsymbol{x}_{i}-\tau_{i}\right) \cdot \boldsymbol{v}_{j}} \\
&=2-\left\|\boldsymbol{x}_{\boldsymbol{i}}-\boldsymbol{\tau}_{\boldsymbol{i}}\right\|^{2}\frac{\left(\boldsymbol{x}_{\boldsymbol{i}}-\boldsymbol{x}_{\boldsymbol{j}}\right) \cdot \boldsymbol{v}_{\boldsymbol{j}}} {x_{0}\left(\boldsymbol{x}_{i}-\tau_{i}\right) \cdot \boldsymbol{v}_{\boldsymbol{j}}} \\
&=2-\left\|\boldsymbol{x}_{\boldsymbol{i}}-\boldsymbol{\tau}_{\boldsymbol{i}}\right\| \cdot \frac{\left(\boldsymbol{x}_{i}-\boldsymbol{x}_{j}\right) \cdot \boldsymbol{v}_{j}}{x_{0}\left\|\boldsymbol{v}_{\boldsymbol{j}}\right\| \cos \theta} 
\end{align}

Therefore, \\
\begin{align}
&\lim _{x_{i} \rightarrow \tau_{i}} \frac{\left(\nabla B_{i x_{i}}\right)^{\mathrm{T}} \cdot v_{j}}{\left(x_{i}-\tau_{i}\right)^{\mathrm{T}} \cdot v_{j}}=2 \\ &\Longrightarrow
\lim _{x_{i} \Longrightarrow \tau_{i}} \frac{\left(x_{i}-\tau_{i}\right)^{\mathrm{T}} \cdot v_{j}}{\left(\nabla B_{i x_{i}}\right)^{T} \cdot v_{j}}=\frac{1}{2} \\
&\Longrightarrow
\lim _{\boldsymbol{x}_{i} \Rightarrow \tau_{i}}\left(1-\frac{2\left(\boldsymbol{x}_{i}-\tau_{i}\right)^{\mathrm{T}} \cdot v_{j}}{x_{0}\left(\nabla B_{i x_{i}}\right)^{\mathrm{T}} \cdot \boldsymbol{v}_{j}}\right) \boldsymbol{v}_{\boldsymbol{j}}=\mathbf{0}
\end{align}

Since 
\begin{equation}
\nabla B_{i x_{i}}\left(\tau_{i}\right)=0,\\ \lim _{x_{i} \Rightarrow \tau_{i}}-k_{1}\left\|\nabla B_{i x_{i}}\right\|^{\alpha-1} \nabla B_{i x_{i}}=0
\end{equation}
\\
which leads to
\begin{multline}
\lim _{\boldsymbol{x}_{i} \rightarrow \tau_{i}}-k_{1}\left\|\nabla B_{i x_{i}}\right\|^{\alpha-1} \nabla B_{i x_{i}}+\\
\left(1-\frac{2\left(\boldsymbol{x}_{i}-\tau_{i}\right)^{\mathrm{T}} \cdot v_{j}}{x_{0}\left(\nabla B_{i x_{i}}\right)^{\mathrm{v}} \cdot v_{j}}\right) \boldsymbol{v}_{\boldsymbol{j}}=\mathbf{0}
\end{multline}

Lemma 2: Time derivative of the barrier function $\dot{B}_{i}$ satisfies:
\begin{equation}
\dot{B}_{i}=-k_{1}\left\|\nabla B_{i x_{i}}\right\|^{\alpha+1}
\end{equation}

Proof: 
\begin{align}
\dot{B}_{i} &=\left(\nabla B_{i x_{i}}\right)^{\mathrm{T}} \cdot \boldsymbol{v}_{\boldsymbol{i}}+\left(\nabla B_{i x_{j}}\right)^{\mathrm{T}} \cdot \boldsymbol{v}_{\boldsymbol{j}} \\
&=\left(\nabla B_{i x_{i}}\right)^{\mathrm{T}} \cdot[-k_{1}\left\|\nabla B_{i x_{i}}\right\|^{\alpha-1} \nabla B_{i x_{i}} \\
&\qquad+\left(1-\frac{2\left(\boldsymbol{x}_{i}-\boldsymbol{\tau}_{i}^{\mathrm{T}} \cdot \boldsymbol{v}_{j}\right.}{x_{0}\left(\nabla B_{i x_{i}}\right)^{\mathrm{T}} \cdot \boldsymbol{v}_{j}}\right)\boldsymbol{v}_{j}]\\
&\qquad+\frac{\left\|\boldsymbol{x}_{\boldsymbol{i}}-\boldsymbol{\tau}_{i}\right\|^{2}}{x_{0}\left\|\boldsymbol{x}_{i}-\boldsymbol{x}_{j}\right\|} \cdot\left(\boldsymbol{x}_{\boldsymbol{i}}-\boldsymbol{x}_{\boldsymbol{j}}\right)^{\mathrm{T}} \cdot \boldsymbol{v}_{\boldsymbol{j}} \\
&=-k_{1}\left\|\nabla B_{i x_{i}}\right\|^{\alpha+1}+2 \cdot \frac{\left(\boldsymbol{x}_{i}-\boldsymbol{\tau}_{i}\right)^{\mathrm{T}} \cdot \boldsymbol{v}_{\boldsymbol{j}}}{x_{0}} \\
&\qquad-\left(\nabla B_{i x_{i}}\right)^{\mathrm{T}} \cdot \boldsymbol{v}_{\boldsymbol{j}} \cdot \frac{2\left(\boldsymbol{x}_{i}-\boldsymbol{\tau}_{i}\right)^{\mathrm{T}} \cdot \boldsymbol{v}_{j}}{x_{0}\left(\nabla B_{i x_{i}}\right)^{\mathrm{T}} \cdot \boldsymbol{v}_{\boldsymbol{j}}} \\
&=-k_{1}\left\|\nabla B_{i x_{i}}\right\|^{\alpha+1}
\end{align}

Lemma 3: In the domain $\mathbf{D}_{0}=\left\{\mathbf{x} \mid\left\|\boldsymbol{x}-\boldsymbol{x}_{\boldsymbol{j}}\right\|>d_{c}\right\},$
\begin{equation}
B(x) \leq \epsilon\left\|x-\tau_{i}\right\|^{2}
\end{equation} \\

Proof:
\begin{align}
\left\|x-x_{j}\right\| \geq d_{c} \Longrightarrow\left\|x-x_{j}\right\|-d_{c} \geq 0 \\
\Longrightarrow\left\|x-x_{j}\right\|-d_{c}+\frac{1}{\epsilon} \geq \frac{1}{\epsilon} \\
\Longrightarrow \frac{1}{\left\|x-x_{j}\right\|-d_{c}+\frac{1}{\epsilon}} \leq \epsilon \\
\Longrightarrow B_{i}=\frac{\left\|x-\tau_{i}\right\|^{2}}{\left\|x-x_{j}\right\|-d_{c}+\frac{1}{\epsilon}} \leq \epsilon\left\|x-\tau_{i}\right\|^{2}
\end{align}

Lemma 4: $\nabla B_{i, x_{i}}$ is non-zero everywhere except the equilibrium point $\tau_{i}$, the point
\begin{equation}
\boldsymbol{x}=\tau_{i}+2 \frac{\left\|\boldsymbol{x}_{j}-\tau_{i}\right\|+d_{c}-\frac{1}{\underline{c}}}{\left\|\boldsymbol{x}_{j}-\boldsymbol{\tau}_{i}\right\|}\left(\boldsymbol{x}_{j}-\tau_{i}\right)
\end{equation}

and the point
\begin{equation}
\boldsymbol{x}=\boldsymbol{x}_{j}+2 \frac{\left\|\boldsymbol{\tau}_{i}-\boldsymbol{x}_{j}\right\|+d_{\varepsilon}-\frac{1}{\alpha}}{\left\|\boldsymbol{\tau}_{i}-\boldsymbol{x}_{j}\right\|}\left(\tau_{i}-\boldsymbol{x}_{j}\right)
\end{equation}

Proof: 
Denote $\nabla B_{i x_{i}}=\frac{\partial B_{i}}{\partial x_{i}}, \nabla B_{i x_{j}}=\frac{\partial B_{i}}{\partial x_{j}}$, and $x_{0}=\left\|\boldsymbol{x}_{i}-\boldsymbol{x}_{j}\right\|-d_{c}+\frac{1}{c}$
\begin{equation}
\nabla B_{i} x_{i}=2 \frac{x-\tau_{i}}{x_{0}}-\frac{\left\|x-\tau_{i}\right\|^{2}}{x_{0}^{2}} \frac{x-o}{\|x-o\|}
\end{equation}

solve the equation gives:
\begin{equation}
\boldsymbol{x}=\boldsymbol{\tau}_{i}+2 \frac{\left\|\boldsymbol{x}_{j}-\tau_{i}\right\|+d_{c}-\frac{1}{\varepsilon}}{\left\|\mathbf{x}_{j}-\tau_{i}\right\|}\left(\boldsymbol{x}_{j}-\tau_{i}\right)
\end{equation}

or
\begin{equation}
\boldsymbol{x}=\boldsymbol{x}_{j}+2 \frac{\left\|\boldsymbol{\tau}_{i}-\boldsymbol{x}_{j}\right\|+d_{c}-\frac{1}{2}}{\left\|\tau_{i}-\boldsymbol{x}_{j}\right\|}\left(\tau_{i}-\boldsymbol{x}_{j}\right)
\end{equation}

Lemma 5: In any closed, compact domain $D \subset \mathbf{R}^{\mathbf{n}}$ containing point $\tau_{i}$ and excluding the region $\bar{D}=\{\boldsymbol{x}\quad|\quad \|\boldsymbol{x}-[\boldsymbol{x}_{j}+2 \frac{\|\tau_{i}-\boldsymbol{x}_{j}\|+d_{c}-\frac{1}{\alpha}}{\\|\boldsymbol{T}_{i}-\boldsymbol{x}_{j}\|}(\tau_{i}-\boldsymbol{x}_{j})]\|<r,$ \\
$\|\boldsymbol{x}-[\boldsymbol{x}_{j}+2 \frac{\|\boldsymbol{\tau}_{i}-\boldsymbol{x}_{j}\|+d_{i}-\frac{1}{2}}{\|\tau_{i}-\boldsymbol{x}_{j}\|}(\tau_{i}-\boldsymbol{x}_{j})]\|<r \}$ where $r$ is an arbitrary small positive number, the gradient of the barrier function satisfies:
\begin{equation}
\left\|\nabla B_{i x_{i}}\right\| \geq c\left\|x-\tau_{i}\right\|
\end{equation}

where $c>0$ \\

Proof: It can be easily verified that $\nabla B_{i x_{i}}\left(\tau_{i}\right)=0 .$ Select $D_{1}=\left\{x \mid\left\|x-\tau_{i}\right\|<\Delta\right\}$, where $\Delta$ is a very small positive number. Choose domain $D=D \backslash D_{1}$ Since  $\mathrm{D}$ doesn't include $\bar{D}$,  $\tilde{D}$ does not include the point as in Lemma $4 .$ Hence, from Lemma 4, at any point $x \in \bar{D}, \nabla B_{i x_{i}} \neq 0$ and since $\bar{D}$ is a closed domain, we can find $c_{1}=\min _{x \in \bar{D}} \frac{\left\|\nabla B_{i x_{i}}\right\|}{\left\|_{x-\tau}\right\|}>0$. Therefore, we have that $\forall x \in \bar{D},\left\|\nabla B_{i x_{i}}\right\| \geq c_{1}\left\|x-\tau_{i}\right\|$
Next, consider $D_{2}=\left\{x \mid\left\|x-\tau_{i}\right\| \leq \Delta\right\}$. In a very small neighborhood of $\tau_{i}$, the Hessian natrix $\nabla^{2} B_{i x_{i}}>0$. Hence, using First-order condition for convexity, we have that $\forall x \in D_{2}$,
\begin{align}
B_{i}\left(\tau_{i}\right) \geq B_{i}+\nabla B_{i x_{i}}^{T}\left(\tau_{i}-x\right) \\
\Longrightarrow 0 \geq B_{i}-\nabla B_{i x_{i}}^{T}\left(x-\tau_{i}\right) \\
\Longrightarrow \nabla B_{i x_{i}}^{T}\left(x-\tau_{i}\right) \geq B_{i}
\end{align}

It is obvious that $B_{i}$ can be bounded as $B_{i} \geq c_{2}\left\|x-\tau_{i}\right\|^{2}$. Also, using Cauchy-Schwartz inequalty, we have that $\nabla B_{i x_{i}}^{T}\left(x-\tau_{i}\right) \leq\left\|\nabla B_{i x_{i}}\right\|\left\|x-\tau_{i}\right\|$ \\

Therefore, 
\begin{align}
\left\|\nabla B_{i x_{i}}\right\|\left\|x-\tau_{i}\right\| \geq \nabla B_{i x_{i}}^{T}\left(x-\tau_{i}\right) \\
\quad \geq B_{i} \geq c_{2}\left\|x-\tau_{i}\right\|^{2} \\
\Longrightarrow\left\|\nabla B_{i x_{i}}\right\| \geq c_{2}\left\|x-\tau_{i}\right\|.
\end{align}

Now we can give the proof of Theorem 1: \\

Proof: From Lemma 4 , we have that $\nabla B_{i x_{i}}=\mathbf{0}$ at the equilibrium point $\tau_{i}$ and at the point $x=\tau_{i}+\mu\left(o-\tau_{i}\right)$ where $\mu$ takes the value as per Lemma 3 . Lets assume that the initial condition is such that $x\left(t_{0}\right)$ doesn't lie in $\bar{D}$ defined as per Lemma 4. Consider the open domain around the goal location $D_{0}$ as defined in Lemma 2 . Define $\mathcal{D}=D_{0} \backslash \bar{D}$ since $\bar{D}$ is a closed domain and $D_{0}$ is open, domain $\mathcal{D}$ is an open domain around the equilibrium $\tau_{i}$.
Choose the candidate Lyapunov function
\begin{equation}
V_{i}=B_{i}
\end{equation}

From Lemma 2 we have: 
\begin{equation}
\dot{V}_{i}=-k_{1}\left\|\nabla B_{i x_{i}}\right\|^{\alpha+1}
\end{equation}

\begin{figure*}[ht]
    \centering
    \includegraphics[width=0.6\textwidth]{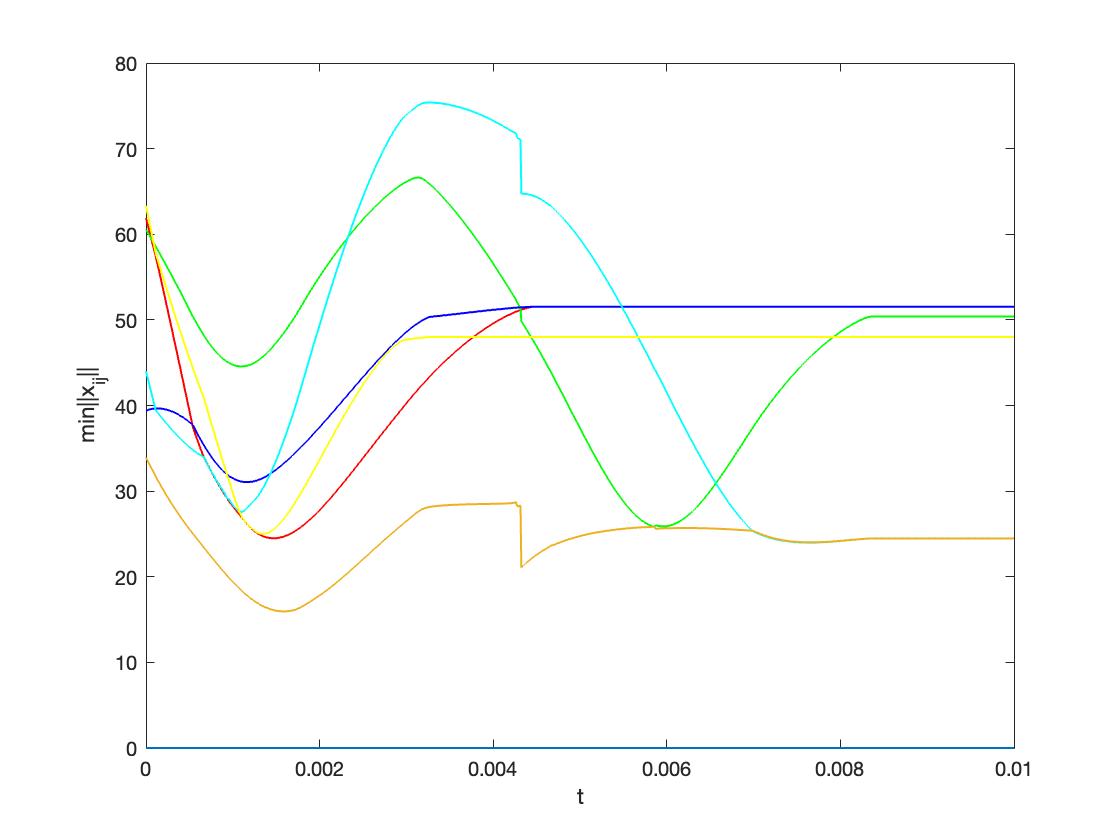}
    \caption{Example 1: Distance between 4 agents.}
    \label{Fig.4}
\end{figure*}

From Lemma 5 we have:
\begin{equation}
\left\|\nabla B_{i x_{i}}\right\| \geq c_{0}\left\|x-\tau_{i}\right\| \Longrightarrow\left\|x-\tau_{i}\right\| \leq \frac{\left\|\nabla B_{i x_{i}}\right\|}{c_{0}}
\end{equation}

From Lemma 3 we have:
\begin{equation}
\begin{gathered}
V_{i}=B_{i} \leq \epsilon\left\|\boldsymbol{x}-\tau_{\boldsymbol{i}}\right\|^{2} \leq \epsilon \cdot \frac{\left\|\nabla B_{i x_{i}}\right\|^{2}}{c_{0}^{2}} \\
\Longrightarrow\left\|\nabla B_{i x_{i}}\right\|^{2} \geq \frac{c_{0}^{2}}{\epsilon} B_{i}
\end{gathered}
\end{equation}

Therefore, we have:
\begin{align}
\dot{V}_{i}&=-k_{1}\left\|\nabla B_{i x_{i}}\right\|^{\alpha+1}\\
&\leq-k_{1} \frac{c_{0}^{\alpha+1}}{\epsilon^{\alpha+1}}\left(B_{i}\right)^{\frac{\alpha+1}{2}}=-k_{1} \frac{c_{0}^{\alpha+1}}{\epsilon^{\alpha+1}}\left(V_{i}\right)^{\frac{\alpha+1}{2}}
\end{align}

If we set $k_{1} \frac{c_{0+1}^{\alpha+1}}{\epsilon^{\frac{\alpha+1}{2}}}=c>0$ and $0<\frac{\alpha+1}{2}=\beta<1$, then
\begin{equation}
\dot{V}_{i} \leq-c V_{i}{ }^{\beta}
\end{equation}

which satisfies the condition of FTS. \\
\section{Numerical Simulations}
\subsection{Example 1: Cases of 4 agents}
To verify the efficacy of the proposed controller, we demonstrated the whole process using MATLAB. The first case involves 4 agents, where the radius of every agent is set to be r = 0.99. The radius of total move area R = 98. We choose $d_c = 2, \epsilon= 10000$ in the barriar function, and $\alpha = \frac{1}{3}$ in the controller. The time interval for the simulation is $10^{-3}s$. The evolution of the inter-agent distances during the entire simulation time is depicted in Fig. 1. The motion of the followers towards their destinations is depicted in Fig. 3. All agents arrive at their destinations within finite time.
\subsection{Example 2: Cases of 20 agents}
 The second case involves 20 agents and the time interval is set to be $2\times10^{-3}s$. Other settings are the same as case 1. The evolution of the inter-agent distances during the entire simulation time is depicted in Fig. 2. The motion of the followers towards their destinations is depicted in Fig. 4. All agents arrive at their destinations within finite time.
 
\begin{figure}[ht]    \includegraphics[width=0.5\textwidth]{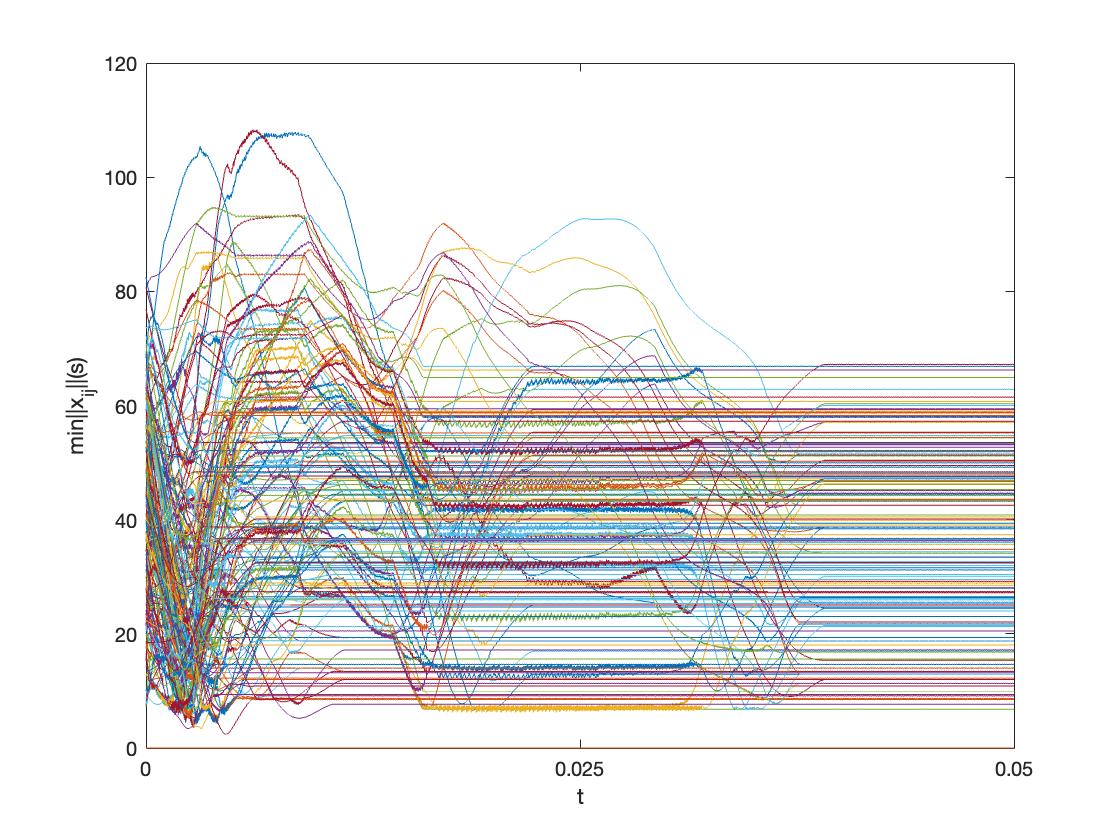}
    \caption{Example 2: Distance between 20 agents}
    \label{1}
\end{figure} 

\begin{figure}
	\centering
	\subfigure[$t=0$]{
		\includegraphics[width=0.3\textwidth,height=0.25\textwidth]{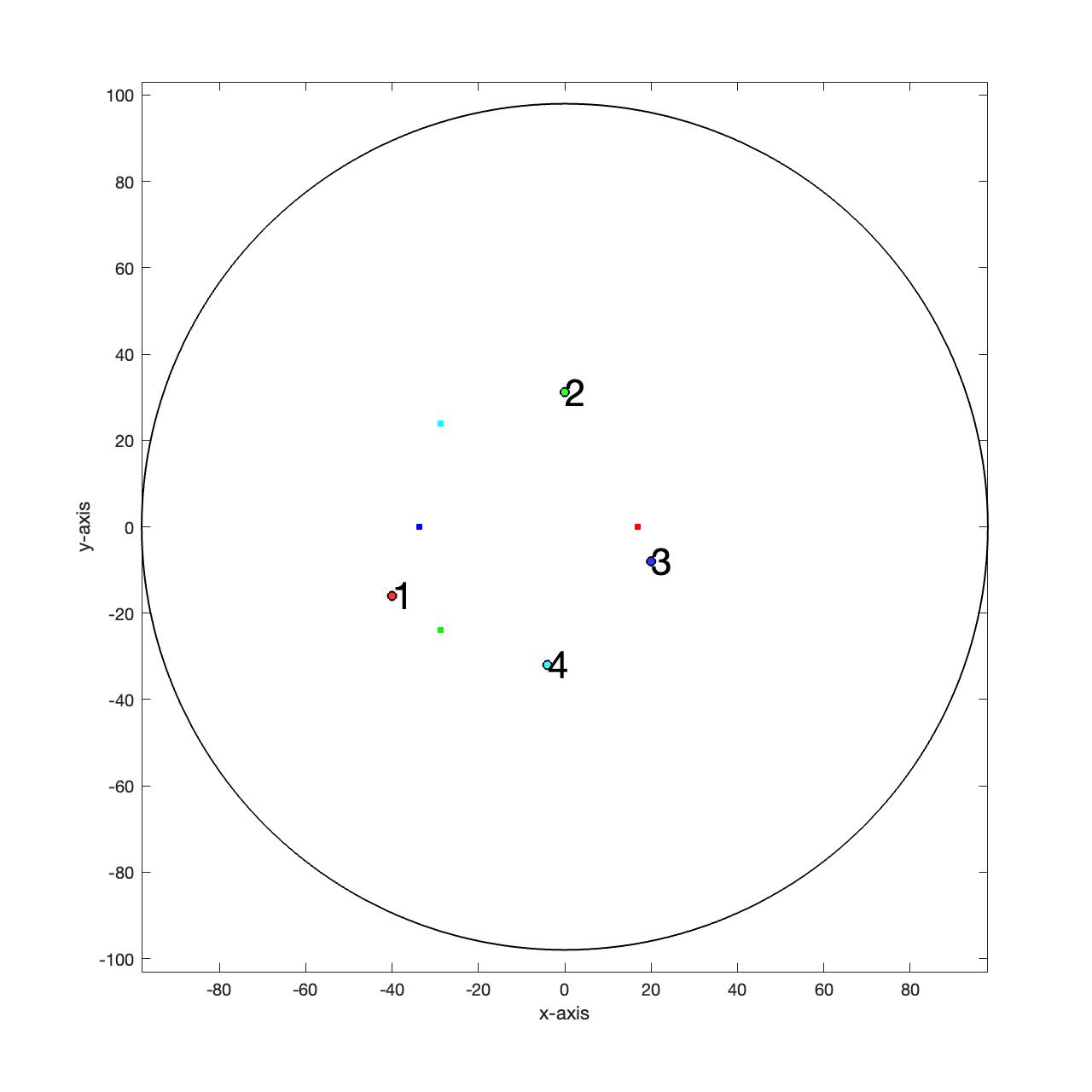} 
		\label{2}
	}
	\subfigure[$t=2\times10^{-3}s$]{
   	 	\includegraphics[width=0.3\textwidth,height=0.25\textwidth]{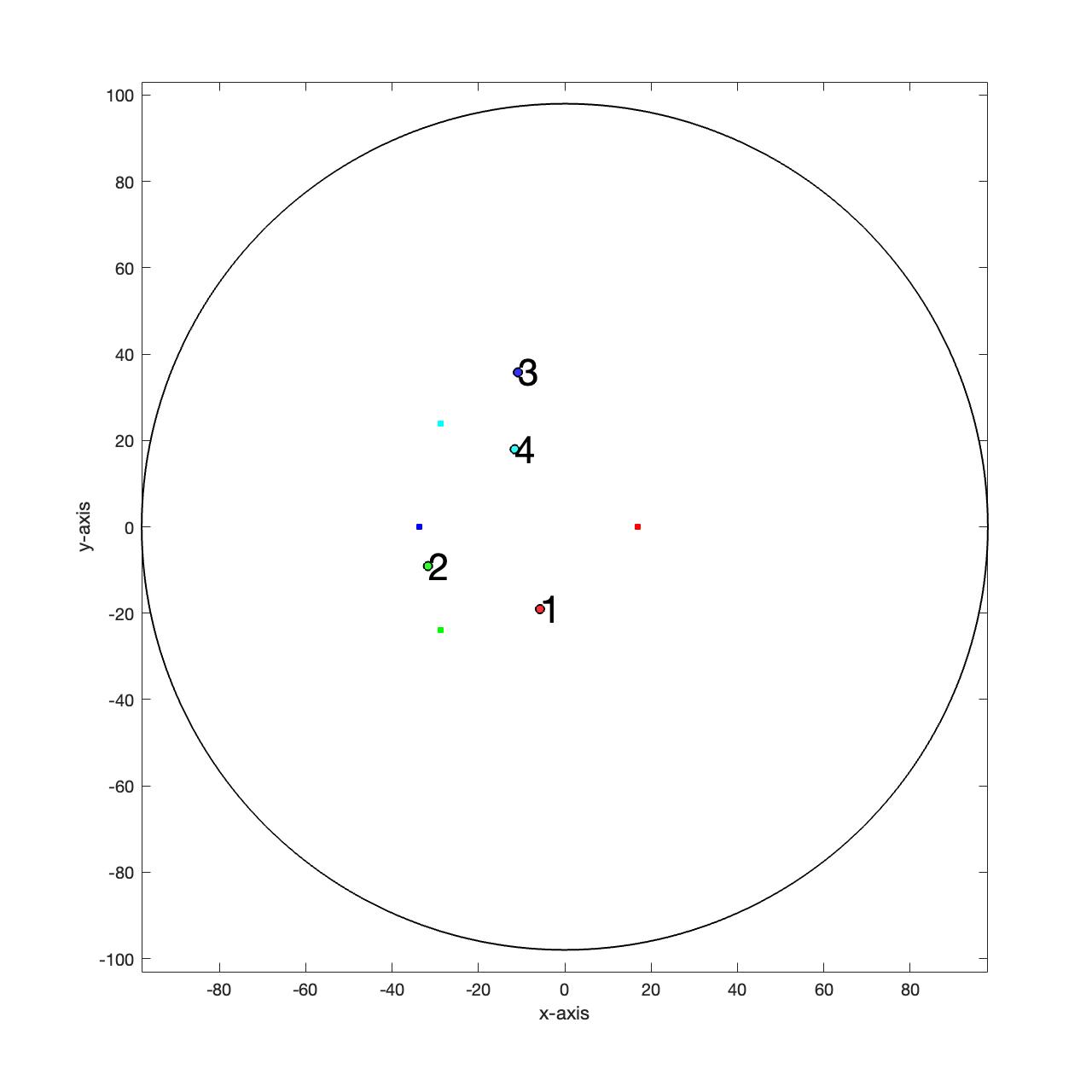}
	\label{3}
	}
	\subfigure[$t=8\times10^{-3}s$]{
			\includegraphics[width=0.3\textwidth,height=0.25\textwidth]{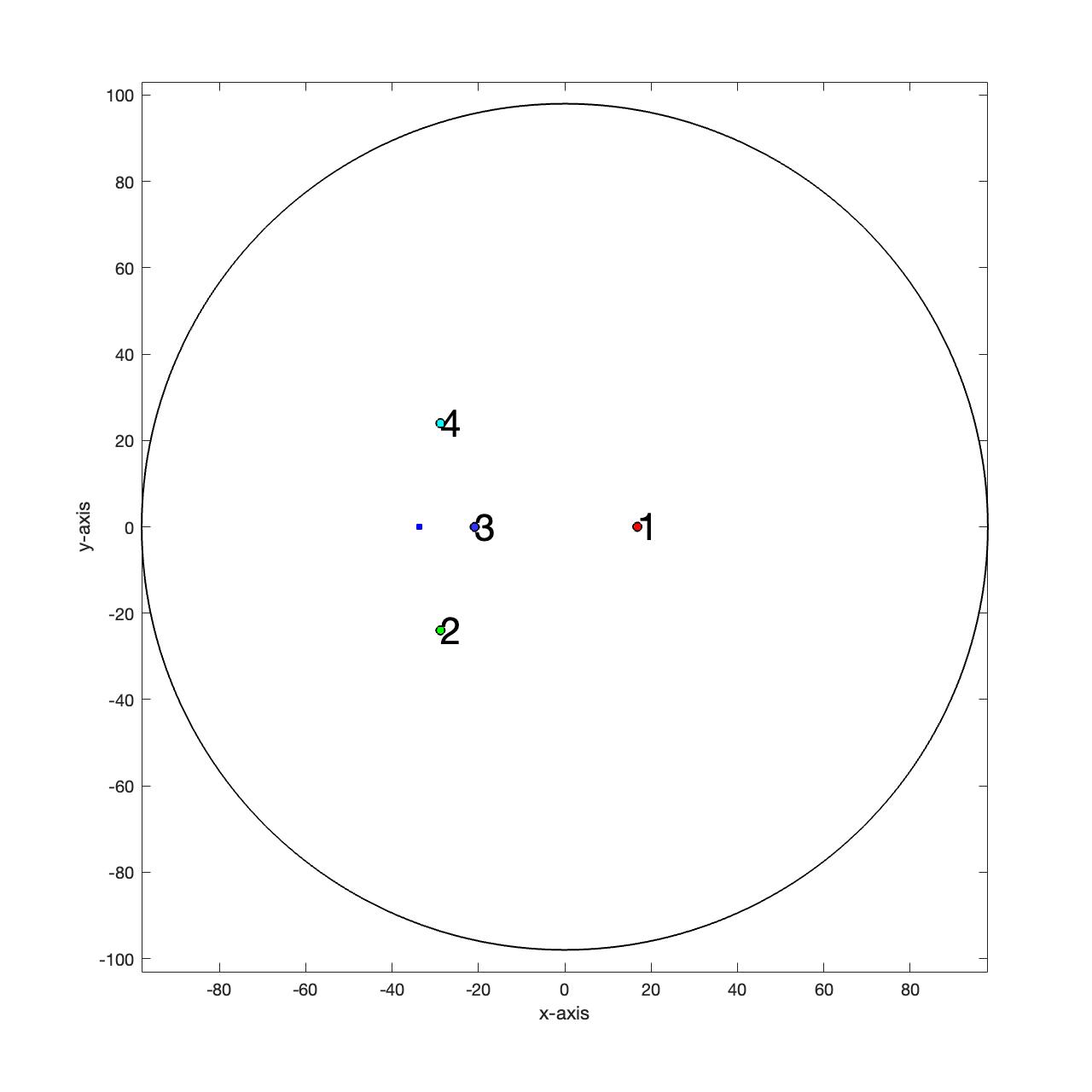} 
		\label{4}
	}
	\subfigure[$t=1\times10^{-2}s$]{
			\includegraphics[width=0.3\textwidth,height=0.25\textwidth]{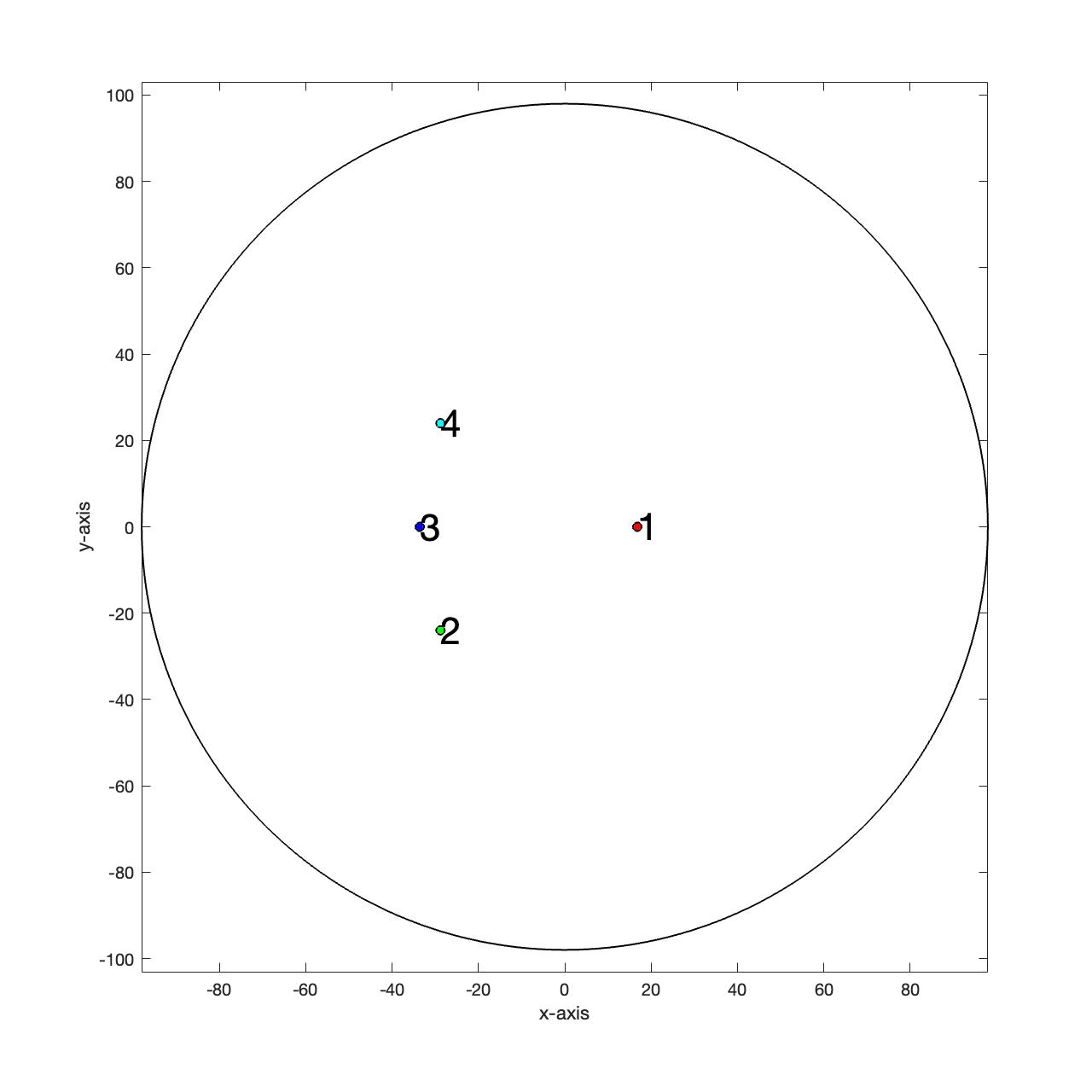} 
		\label{5}
	}
	\caption{States of four agents}
\end{figure}

\begin{figure}
	\subfigure[$t=0$]{
		\includegraphics[width=0.3\textwidth,height=0.25\textwidth]{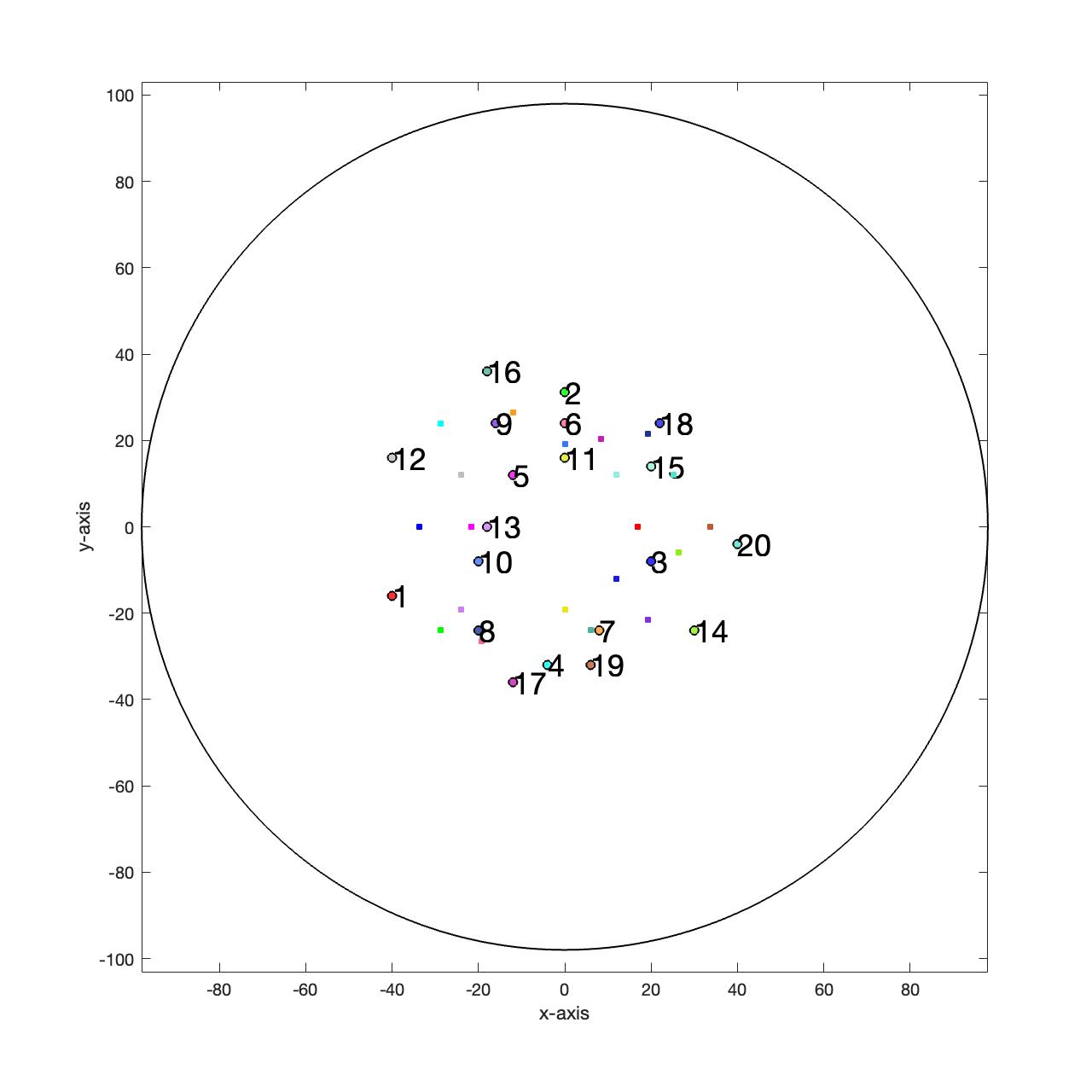} 
		\label{6}
	}
	\subfigure[$t=1\times10^{-3}s$]{
   	 	\includegraphics[width=0.3\textwidth,height=0.25\textwidth]{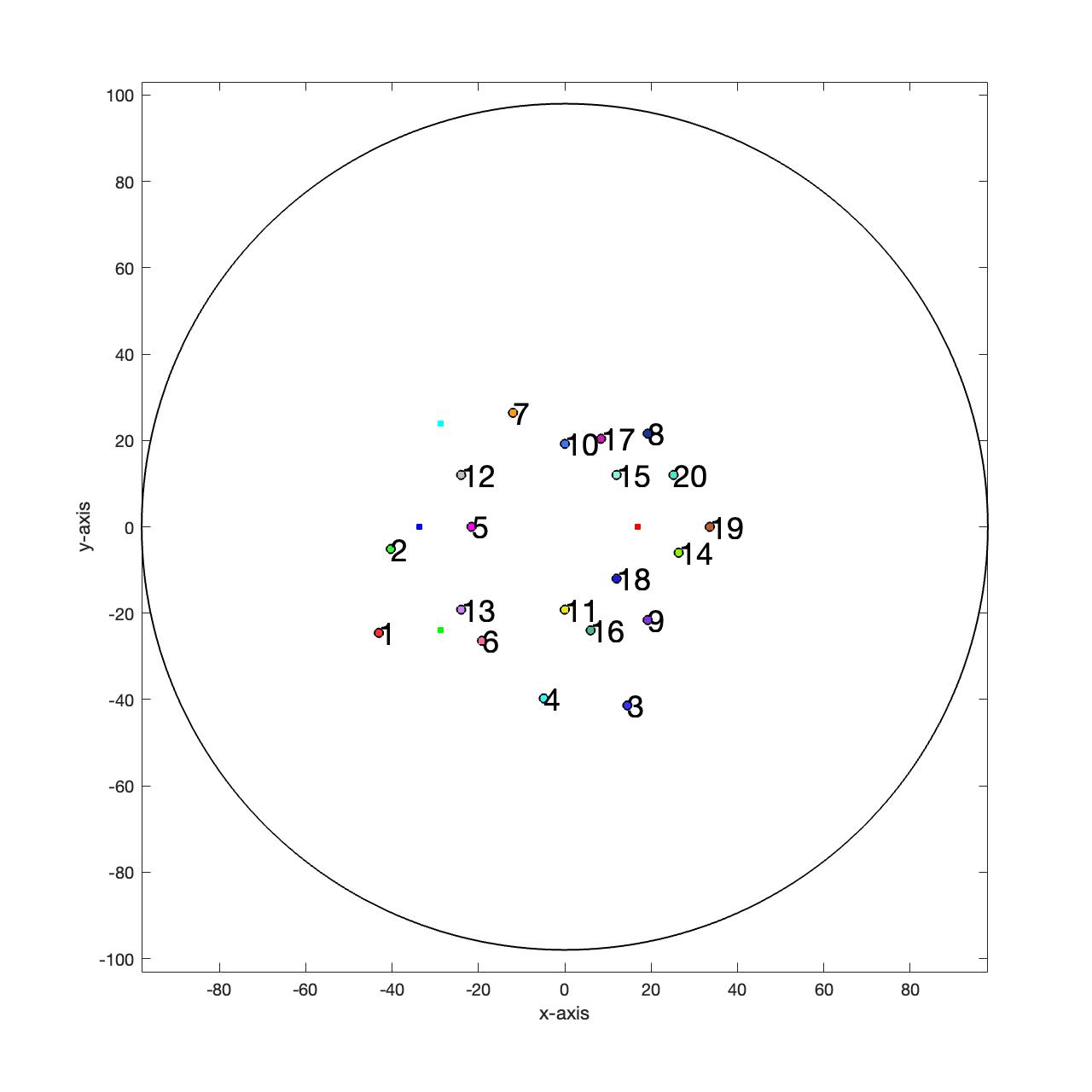}
	\label{7}
	}
	\subfigure[$t=1\times10^{-2}s$]{
        \includegraphics[width=0.3\textwidth,height=0.25\textwidth]{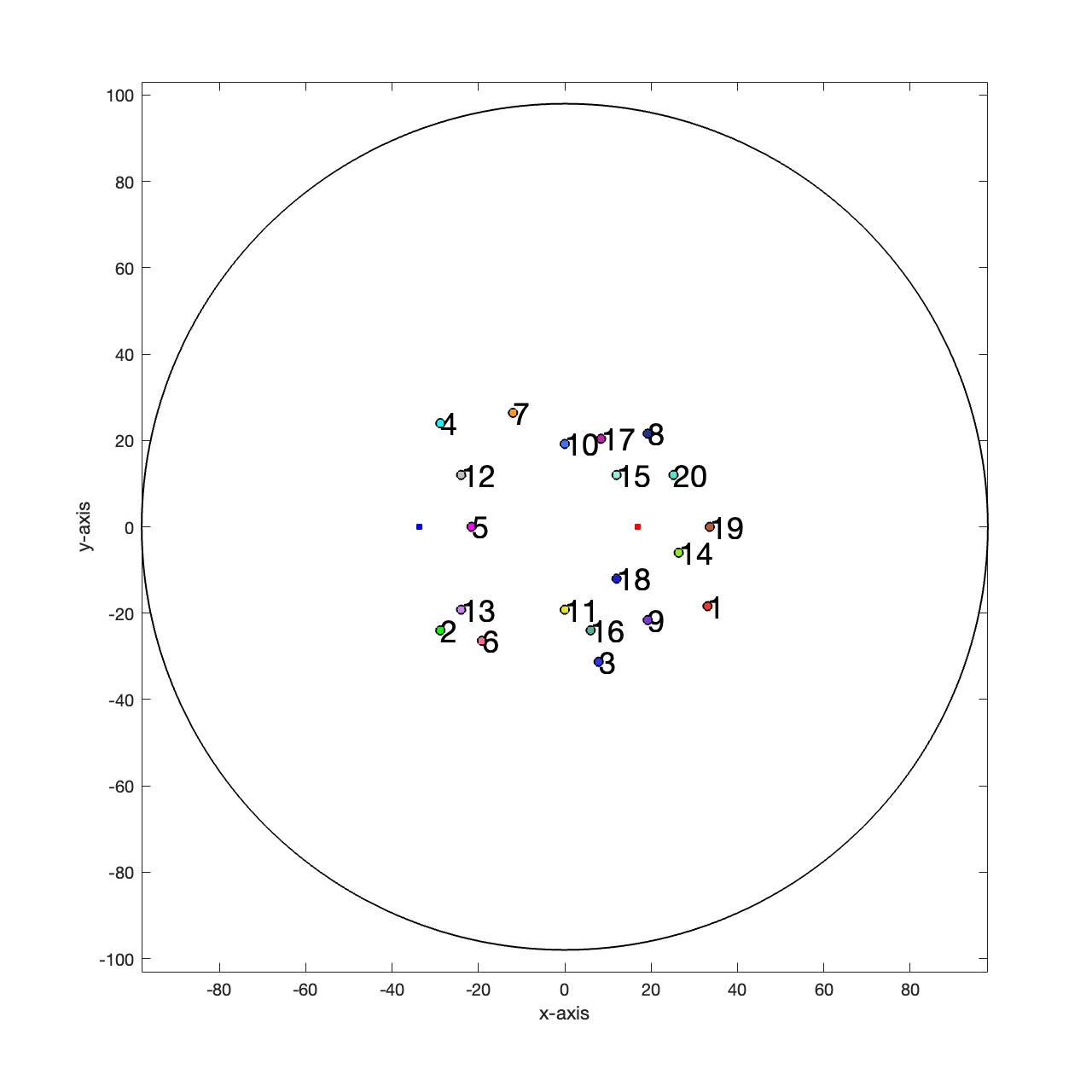} 
	\label{8}
	}
	\subfigure[$t=5\times10^{-2}s$]{
   	 	\includegraphics[width=0.3\textwidth,height=0.25\textwidth]{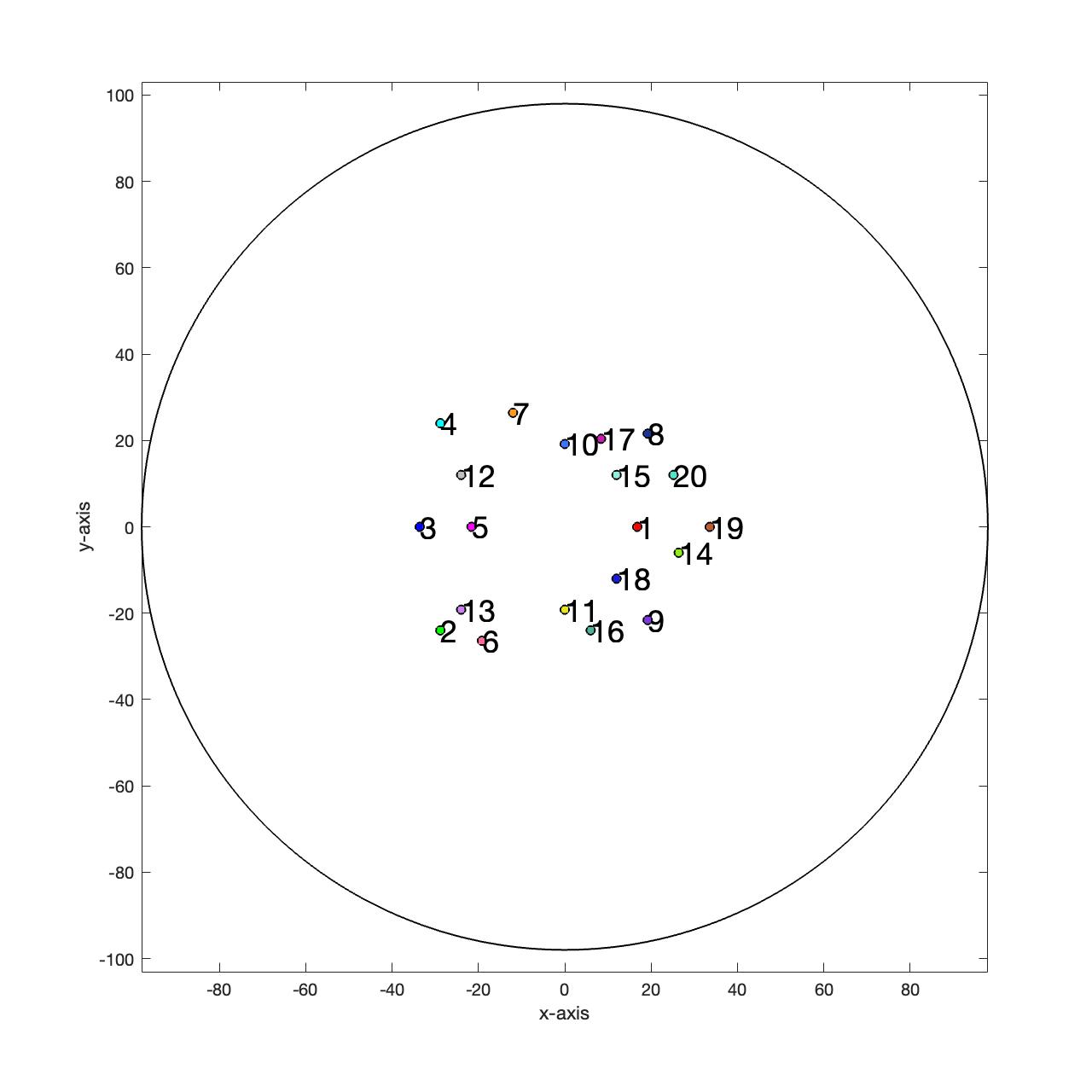}
	\label{9}
	}
	\caption{States of twenty agents}
\end{figure}

\section{Conclusion}
 In this paper, we propose a finite-time distributed controller with considering collision avoidance for multi-agent systems. Our main contribution lies in three aspects: First, a new type of Lyapunov barrier function is   proposed, whose derivative satisfy specific properties such that its gradient is non-zero everywhere except the equilibrium point; Then, a distributed controller for multi-agent systems with considering collision avoidance is provided by using the developed Lyapunov barrier function; Finally, the proposed method can deal with the situations where unsafe region or boundaries of safe region exists, providing a more practical implementation environment for multi-agent systems.\\

This paper considers a simple first-order dynamical multi-agent system, while sophisticated models could be investigated with a more practical setting up. Therefore, our future efforts will be devoted to the follows: A second-order dynamic model of nonholonomic mobile agents will be considered with applications in real wheeled swarm robots \cite{b10,dimarogonas07tac}; We will consider more practical constraints in the agents' sensing and communication, like the problem of connectivity maintainance and preservation \cite{b12,kan12tac,han2016estimating,han2014tcas1,han12tii}; Design of robust distributed controller is a promising way to cope with the disturbance in dynamics and communications, and parametric Lyapunov-like barrier function might be exploited \cite{han19tac}. Learning-based methods by using Lyapunov-barrier functions will also be attempted \cite{huang21ccdc}.


\begin{thebibliography}{00}



\bibitem{pecora90prl} L.~M. Pecora and T.~L. Carroll, ``Synchronization in chaotic system,''
{\em Physical Review Letters}, 64(8):1146--1152, 1990.

\bibitem{jlu04cas1} J.~L\"{u}, G.~Chen, X.~Yu, and H.~Leung, ``Design and analysis of multiscroll chaotic attractors from saturated function series,''
{\em IEEE Transactions on Circuits and Systems-I},51(12):2476--2490,2004.
\bibitem{zli10} Z.~Li, Z.~Duan, G.~Chen, and L.~Huang,
``Consensus of multi-agent systems and synchronization of complex networks: A unified viewpoint'',
{\em IEEE Transactions on Circuits and Systems-I}, 57(1):213--224,
\bibitem {panagou16tac}  D. Panagou and D. M. Stipanovi{\'c},  and P. Voulgaris, ``Distributed coordination control for multi-robot networks using {L}yapunov-like barrier functions'',
{\em IEEE Transactions on Automatic Control}, vol. 61, no. 3, pp. 617--632, 2016.
\bibitem {han19tac}  D. Han and D. Panagou, ``Robust multitask formation control via parametric {L}yapunov-like barrier functions'', 
{\em IEEE Transactions on Automatic Control}, vol. 64, no. 11, pp. 4439--4453, 2019.
\bibitem{b1} A. Jadbabaie, J. Lin, and A. S. Morse, ''Coordination of groups of mobile autonomous agents using nearest neighbor rule'', 
{\em IEEE Transactions on Automatic Control}, vol. 48, no. 6, pp. 988–1001, Jun. 2003
\bibitem{b2} H. G. Tanner, “Flocking with obstacle avoidance in switching networks of interconnected vehicles,” in 
{\em Proceedings of IEEE International Conference on Robotics and Automation}, 2004, pp. 3006–3011.
\bibitem{b3} R. Olfati-Saber, “Flocking for multi-agent dynamic systems: Algorithms and theory'', {\em IEEE Transactions on Automatic Control}, vol. 51, no. 3, pp. 401–420, Mar. 2006.
\bibitem{b4} H. G. Tanner, A. Jadbabaie, and G. J. Pappas, “Flocking in fixed and switching networks,” 
{\em IEEE Transactions on Automatic Control}, vol. 52, no. 5, pp. 863–868, May 2007.
\bibitem{b5} B. Sharma, J. Vanualailai, and U. Chand, “Flocking of multi-agents in constrained environments,” 
{\em European Journal of Pure and Applied Mathematics.}, vol. 2, no. 3, pp. 401–425, 2009.
\bibitem{b6} H. Su, X. Wang, and Z. Lin, “Flocking of multi-agents with a vir- tual leader,” 
{\em IEEE Transactions on Automatic Control}, vol. 54, no. 2, pp. 293–307, Feb. 2009.
\bibitem{b7} M. Ji and M. Egerstedt, “Distributed coordination control of multiagent systems while preserving connectedness,”
{\em IEEE Transactions on Robotics}, vol. 23, no. 4, pp. 693–703, Aug. 2007.
\bibitem{b8} R. Olfati-Saber, J. A. Fax, and R. M. Murray, “Consensus and cooperation in networked multi-agent systems,” in {\em Proceedings of IEEE}, vol. 95, no. 1, pp. 215–233, Jan. 2007.
\bibitem{b9} D. V. Dimarogonas and K. J. Kyriakopoulos, “A connection between formation infeasibility and velocity alignment in kinematic multi-agent systems,” 
{\em Automatica}, vol. 44, no. 10, pp. 2648–2654, Oct. 2008.
\bibitem{b10} S. Mastellone, D. M. Stipanovic, C. R. Graunke, K. A. Intlekofer, and M. W. Spong,  `` Formation control and collision avoidance for multi-agent non-holonomic systems: Theory and experiments,” 
{\em The International Journal of Robotics Research}, vol. 27, no. 1, pp. 107–126, Jan. 2008.
\bibitem{b11}D. V. Dimarogonas and K. J. Kyriakopoulos, “Connectedness preserving distributed swarm aggregation for multiple kinematic robots,” 
{\em IEEE Transactions on Robotics}, vol. 24, no. 5, pp. 1213–1223, Oct. 2008.
\bibitem{b12}M. M. Zavlanos, M. B. Egerstedt, and G. J. Pappas, “Graph-theoretic connectivity control of mobile robot networks,” in 
{\em Proceedings of IEEE}, vol. 99, no. 9, pp. 1525–1540, Sep. 2011.
\bibitem{b13}Hernández-Martínez, E. G., and Aranda-Bricaire, E.,  ''Convergence and collision avoidance in formation control: A survey of the artificial potential functions approach'',  {\em Rijeka, Croatia: INTECH Open Access Publisher}, pp. 105-126.
\bibitem{b14}Bhat, S. P. , Bernstein, D. S., ''Finite-time stability of continuous autonomous systems'',
{\em SIAM Journal on Control and Optimization}, $38(3), 751-766$.
\bibitem{b15}L. Wang and F. Xiao, "Finite-Time Consensus Problems for Networks of Dynamic Agents,"
{\em IEEE Transactions on Automatic Control}, vol. 55, no. 4, pp. 950-955, April 2010
\bibitem{b16}Shang, Yilun, ''Finite-time consensus for multi-agent systems with fixed topologies'', {\em International Journal of Systems Science}, 43. 1-8. 10.1080/00207721.2010.517857. 
\bibitem{b17}Zuo, Z. , Tie, L., ''A new class of finite-time nonlinear consensus protocols for multi-agent systems'', {\em International Journal of Control}, $87(2), 363-370 .$
\bibitem{b18}Flores-Resendiz, J. F. , Aranda-Bricaire, E. , Gonzalez-Sierra, J. , Santiaguillo-Salinas, J, ''Finite-time formation control without collisions for multiagent systems with communication graphs composed of cyclic paths'',
{\em Mathematical Problems in Engineering}, $2015,(2015-5-18)$, $2015(\mathrm{PT} .8), 1-17 .$
\bibitem{b19}Garg, K., Panagou, D., ''New results on finite-time stability: Geometric conditions and finite-time controllers.'' in 
{\em 2018 Annual American Control Conference (ACC)} (pp. 442-447).
\bibitem{dimarogonas07tac} 
D.~V.~Dimarogonas and K.~J.~Kyriakopoulos, ``On the rendezvous problem for multiple nonholonomic agents,'' 
{\em IEEE Transactions on Automatic Control}, vol.~52, no.~5, pp.~916--922, 2007.

\bibitem{kan12tac} 
Z.~Kan  and A.~P.~Dani and J.~M.~Shea and W.~E.~Dixon, 
``Network connectivity preserving formation stabilization and obstacle avoidance via a decentralized controller,'' 
{\em IEEE Transactions on Automatic Control}, vol.~57, no.~7, pp.~1827--1832, 2012.

\bibitem{han2016estimating}
D.~Han and M.~Althoff, 
``On estimating the robust domain of attraction for uncertain non-polynomial systems: {An} {LMI} approach,'' 
in {\em Proceedings of the Conference on Decision and Control}, pp.~2176--2183, 2016.

\bibitem{han2014tcas1}
D.~Han and G.~Chesi, 
``Robust synchronization via homogeneous parameter-dependent polynomial contraction matrix,'' 
{\em IEEE Transactions on Circuits and Systems I: Regular Papers}, vol.~61, no.~10, pp.~2931--2940, 2014.

\bibitem{han12tii}
D.~Han, G.~Chesi, and Y.~S. Hung, 
``Robust consensus for a class of uncertain multi-agent dynamical systems,'' 
{\em IEEE Transactions on Industrial Informatics}, vol.~9, no.~1, pp.~306--312, 2012.

\bibitem{huang21ccdc}
H.~Huang and D.~Han, ``On Estimating the Probabilistic Region of Attraction for Partially Unknown Nonlinear Systems: An Sum-of-Squares Approach,'' 
in {\em Proceedings of the Chinese Control and Decision Conference}, 2022, Accepted.

\end{thebibliography}
\end{document}